\documentclass[12pt,english,aip,amsmath,amssymb,reprint,floatfix]{revtex4-1}
\usepackage[T1]{fontenc}
\usepackage[latin9]{inputenc}
\usepackage{textcomp}
\usepackage{multirow}
\usepackage{amsmath}
\usepackage{graphicx}
\usepackage{esint}

\makeatletter

\providecommand{\tabularnewline}{\\}

\usepackage{array}
\usepackage{enumitem}
\usepackage{afterpage}
\usepackage{caption}
\usepackage{multirow}

\usepackage{babel}

\makeatother

\usepackage{babel}
\begin{document}
\title{Tuning the range separation parameter in periodic systems}
\author{Wenfei Li}
\affiliation{Department of Chemistry and Biochemistry, University of California,
Los Angeles, California 90095, USA}
\author{Vojtech Vlcek}
\thanks{Current address: Department of Chemistry and Biochemistry, University
of California, Santa Barbara, California 93106, USA}
\affiliation{Fritz Haber Center for Molecular Dynamics, Institute of Chemistry,
The Hebrew University of Jerusalem, Jerusalem 91904, Israel}
\author{Helen Eisenberg}
\affiliation{Fritz Haber Center for Molecular Dynamics, Institute of Chemistry,
The Hebrew University of Jerusalem, Jerusalem 91904, Israel}
\author{Eran Rabani}
\affiliation{Department of Chemistry, University of California, and Materials Sciences
Division, Lawrence Berkeley National Laboratory, Berkeley, California
94720, USA}
\affiliation{The Raymond and Beverly Sackler Center for Computational Molecular
and Materials Science, Tel Aviv University, Tel Aviv 69978, Israel}
\author{Roi Baer}
\affiliation{Fritz Haber Center for Molecular Dynamics, Institute of Chemistry,
The Hebrew University of Jerusalem, Jerusalem 91904, Israel}
\author{Daniel Neuhauser}
\affiliation{Department of Chemistry and Biochemistry, University of California,
Los Angeles, California 90095, USA}
\begin{abstract}
Kohn-Sham DFT with optimally tuned range-separated hybrid (RSH) functionals
provides accurate and nonempirical fundamental gaps for a wide variety
of finite-size systems. The standard tuning procedure relies on calculation
of total energies of charged systems and thus cannot be applied to
periodic solids. Here, we develop a framework for tuning the range
separation parameter that can be used for periodic and open boundary
conditions. The basic idea is to choose the range parameter that results
in a stationary point where the fundamental gap obtained by RSH matches
the gap obtained from a $G_{0}W_{0}$ over RSH calculation. The proposed
framework is therefore analogous to eigenvalue self-consistent GW
(scGW). We assess the method for various solids and obtain very good
agreement with scGW results.
\end{abstract}
\maketitle
Due to its high accuracy and low computational cost, Kohn-Sham DFT
(KS-DFT) is one of the most prevalent tools for probing the electronic
structure of both molecular and periodic systems.\cite{Borpuzari2017}
However, KS-DFT with local and semilocal (LDA and GGA) functionals
often severely underestimate the fundamental band gap.\cite{Cohen2011,Faber2014,Borlido}
Several alternative frameworks have been developed to tackle the problem.
One is the GW approximation, where the quasiparticle excitation energies
are obtained by solving a Dyson equation. In practice, calculations
are typically carried out with the $G_{0}W_{0}$ approximation, where
the self energy is applied as a perturbative correction to KS-DFT
orbital energies. For many systems this approach provides quasiparticle
gaps that are in good agreement with experimental band gaps.\cite{Faber2014,Fabien2012,Reining2017,vojtech1,bruneval2016molgw}

Another route is the generalized Kohn-Sham DFT (GKS-DFT) method where
instead of a local exchange-correlation potential, the effective Hamiltonian
is non-local. DFT with hybrid functionals, in either the original
fractional exchange (e.g., B3LYP\cite{becke1993new,lee1988development})
or range-separated flavor, are all part of the GKS framework.\cite{Baer2009,Cohen2011,Baer2010}

In hybrid functionals, certain fraction of Fock exchange is incorporated,
and this choice can be justified by considering two facts. First,
in semilocal approximations, due to the existence of self interaction
the exchange-correlation functional does not have the correct asymptotic
form, while in Hartree-Fock theory the one-particle self interaction
is eliminated through the balance between the Hartree and Fock exchange
terms. Therefore, inclusion of Fock exchange helps achieving the desired
asymptotic behavior of exchange-correlation functionals.\cite{Baer2005,Baer2010}

A second, related, aspect of the self-interaction problem is that
for the exact exchange-correlation functional the total energy curve,
as a function of particle number, should be composed of line segments
joining the energies at integer electron numbers.\cite{perdew1982density,Cohen2007,Cohen2011,Zheng2011,Sanchez2008}
However, DFT with local and semilocal approximations is convex at
fractional charge, while Hartree-Fock is concave. Therefore, by incorporating
Fock exchange, hybrid functionals provide a way of enforcing piecewise
linearity. In fact, studies have shown that optimally tuned range
separated hybrid (OT-RSH) functionals produce total energy curves
that are almost piecewise linear.\cite{RA2012,Korzdorfer2014,Autschbach2014}\\

In OT-RSH the Coulomb interaction between electrons is separated into
short-range and long-range parts:\cite{Baer2010} $\frac{1}{r}=\frac{1-{\rm erf}(\gamma r)}{r}+\frac{{\rm erf}(\gamma r)}{r}$.
The short range part is then approximated using local or semilocal
approximations, which preseve the cancellation of errors between the
exchange and the correlation functional. In long-range exchange functionals,
the long-range part is calculated with Fock exchange to offset the
self-interaction error and enforce the correct long-range asymptotice
behavior of the functional. The range-separation parameter $\gamma$
is chosen to maintain a balance between the long-range and short-range
exchange, and $\gamma^{-1}$ is an effective screening length.

With this partition, the overall exchange-correlation energy becomes
$E_{XC}=E_{C}+E_{F_{X}}^{l}+E_{X}^{s}$, where $E_{F_{X}}^{l}$, $E_{X}^{s}$
are the long range Fock exchange and short-range local/semilocal exchange,
and the superscripts \textquotedbl l\textquotedbl{} and \textquotedbl s\textquotedbl{}
refer to long-range and short-range respectively. The action of the
exchange-correlation part of the Hamiltonian is then:

\begin{align}
\hat{V}_{XC}\psi(r) & =\hat{K}^{l}\psi(r)+[v_{C}(r)+v_{X}^{s}(r)]\psi(r)\nonumber \\
 & =-\int dr'u^{l}(|r-r'|)\rho(r,r')\psi(r')+v_{XC}^{s}(r)\psi(r)\label{eq:lrex}
\end{align}
where $\rho(r,r')$ is the density matrix of the system, $u^{l}(r)=\frac{{\rm erf}\left(\gamma|r|\right)}{|r|}$
is the long-range part of the Coulomb interaction, while $v_{XC}^{s}$
is the short range exchange-correlation potential.

The one unknown is then the range-separation parameter. For finite
sized systems, a self-consistent optimal tuning procedure chooses
$\gamma$ to ensure that Koopman's theorem is obeyed\cite{Autschbach2014,Baer2010,Korzdorfer2014,RA2011,RA2012,kronik2018},
i.e., to minimize the target function:

\begin{equation}
J(\gamma)=|\epsilon_{H}^{\gamma}+IP^{\gamma}|,
\end{equation}
where $\epsilon_{H}^{\gamma}$ is the HOMO energy of the electron
system, and the ionization potential is given by $IP^{\gamma}\equiv[E^{\gamma}(N)-E^{\gamma}(N-\delta)]/\delta$,
where $N$ is the number of electrons in the neutral system, and $\delta$
is a small fractional charge. Optimally tuned RSH (OT-RSH) functionals
have been applied to study various molecular systems and nanocrystals,\cite{cite1}
yielding band gaps in good agreements with GW and/or experimental
results.\cite{stochasticrsh,RA2012,RA2011,Luftner2014,Egger2014,vojtech2}
However, this procedure is not applicable to periodic solids, where
total energy calculations of charged systems are problematic.

There are several ways of obtaining $\gamma$ for solids. For molecular
solids, satisfactory results can be obtained with $\gamma$ tuned
for isolated molecules.\cite{Manna2018} For solids in general, various
attempts were made to connect $\gamma$ with the optical dielectric
constant $\epsilon_{\infty}$.\cite{Baer2009,dielectric1,dielectric2,brawand2016generalization,marques2011density}

Here we devise instead an approach for systematically tuning the range-separation
parameter for periodic systems solely based on first principle calculations.
The idea is to perform RSH calculations as well as $G_{0}W_{0}$ calculations
with RSH as starting points. Two sets of band gaps will then be obtained,
each being functions of the parameter $\gamma$. The optimal $\gamma$
is determined such that the two gaps agree with each other.

We emphasize that this proposed techinque is not just a modified $G_{0}W_{0}$,
but is closely related to the self-consistent GW (scGW) method. In
previous work,\cite{evgw,vojtech1} we have shown that perhaps the
simplest self-consistent GW method is ev-sc$GW_{0}$ with scissors
operator; namely, self-consistently, updating the $G$ operator through
a scissors shift of the occupied vs. virtuals states; this amounts
to repeatedly writing:

\[
\text{\ensuremath{\hat{\Sigma}(t)\propto\hat{\Sigma}(t)e^{-i\Delta\theta(t)t}},}
\]
where we introduce the time domain self-energy $\hat{\Sigma}(t)$,
$\Delta$ is the difference between the quasiparticle band gap and
the DFT band gap, and $\theta(t)$ is the Heaveside step function.
We demonstrated that such a self-cosistent procedure can open up the
fundamental band gap and improve the accuracy of calculated gaps over
one-shot $G_{0}W_{0}$.\cite{evgw,vojtech3} The current approach
for finding $\gamma$ gives $\Delta=0$, and thus amounts to finding
a stationary point for this self-consistent procedure.

In the following discussions, we will refer to this tuning procedure
as OT-GW/RSH.\\

OT-GW/RSH can be implemented with any conventional GW code, such as
VASP.\cite{vasp1,vasp2,vasp3,vasp4} Since many of the OT-GW/RSH applications
are envisioned to eventually take place for large (potentially disordered)
systems, we have also studied here the use of the method with our
recent linear-scaling stochastic GW (sGW) approach, which has been
succesfully applied to systems of $10,000$ electrons and more.\cite{sgw4,sgw3,sgw2,sgw1}
To carry out sGW calculations with RSH, we also applied here a stochastic
method for applying and propagating long-range Fock exchange.\cite{stochasticrsh,bse}

In sGW, the quasiparticle energy in $G_{0}W_{0}$ formulation is obtained
as first-order perturbation to the Kohn-Sham orbital energies:

\begin{equation}
\epsilon_{QP}=\epsilon_{KS}+\langle\phi_{F}|\hat{\Sigma}_{P}(\epsilon_{QP})+\hat{K}-\hat{V}_{XC}|\phi_{F}\rangle\label{eq:eqp}
\end{equation}
where $\phi_{F}(r)$ is typically the HOMO or LUMO, $\hat{\Sigma}_{P}$
is the dynamical polarization self energy, $\hat{K}$ is the full
Fock exchange operator, and $\hat{V}_{XC}$ is the exchange-correlation
part of the GKS-Hamiltonian as in equation (\ref{eq:lrex}). In sGW,
we calculate the matrix elements of $\hat{\Sigma}_{P}$ in the time
domain. Detailed accounts of the sGW method are found in previous
works.\cite{sgw1,sgw2,sgw3,sgw4}

One important issue is that the Generalized Kohn-Sham Hamiltonian
$\hat{h}$ contains the long-range exchange potential, which depends
on the density matrix: $\rho(r,r')=\sum_{i,occ}\phi_{i}(r)\phi_{i}^{*}(r')$.
For large systems, the number of occupied orbitals is large, making
the application of long-range exchange computationally demanding.
To solve this problem, we implemented stochastic long-range Fock exchange
in the sGW code, as done recently.\cite{vlcek2019stochastic} Detailed
explanation of the method can be found in references.\cite{stochasticrsh,vlcek2019stochastic,bse}
In short, we use for the purpose of stochastic exchange a total number
of $N_{\zeta}$ stochastic orbitals, each being a linear combination
of occupied states that are obtained by a low-band-pass flter of a
white-noise function, $|\zeta\rangle=\sqrt{\Theta(\mu-\hat{h})}|\zeta_{0}\rangle$.
Here, we introduced the chemical potential $\mu$, the white noise
function is chosen as $\zeta_{0}(r)\propto\pm1$, and the application
of $\Theta$ operator is carried out by a Chebyshev expansion. The
density matrix is then approximated by:

\begin{equation}
\rho(r,r')\approx[\zeta(r)\zeta^{*}(r')].
\end{equation}

Further, the long-range Coulomb potential is approximated as $u^{l}(|r-r'|)=[\chi(r)\chi^{*}(r')]$,
where $\chi(r)$ is constructed by Fourier transforming a stochastic
combination of the square root of the Fourier components of the long-range
potential, $\sqrt{\tilde{u}^{l}(k)}$.\cite{stochasticrsh} These
two random representations make the long range exchange operator a
sum of separable terms. Then, the action of the long-range exchange
operator becomes:

\begin{align}
\hat{K}^{l}\psi & =-[\zeta(r)\chi(r)\int dr'\zeta^{*}(r')\chi^{*}(r')\psi(r')]\\
 & =-\frac{1}{N_{\zeta}}\sum_{\zeta}\zeta(r)\chi(r)\int dr'\zeta^{*}(r')\chi^{*}(r')\psi(r')
\end{align}

We note that a part of the GW calculation, the action of the short-time
$e^{-i\hat{K}^{l}dt}$ is required; a one-term Taylor expansion is
used, in conjunction with re-normalization of the orbitals after the
short-time propagator is applied. Finally, in calculating the final
quaiparticle energy according to equation (\ref{eq:eqp}), we note
that the difference $\hat{K}-\hat{V}_{XC}$ involves the term $\hat{K}^{s}=\hat{K}-\hat{K}^{l}$.
In our code, this term is calculated in a similar manner to $\hat{K}^{l}$,
but now using the Fourier transofrm of the short range potential.

\section{Results and discussions}

We assessed the tuning procedure for several solids using VASP. For
two of the systems, we additionally performed sGW calculations. As
an illustration of the tuning procedure, we plot the fundamental gaps
calculated from sGW/RSH and RSH, as functions of the range-separation
parameter $\gamma$ for a LiF $5\times5\times5$ supercell.

\begin{figure}
\begin{centering}
\includegraphics[width=8cm]{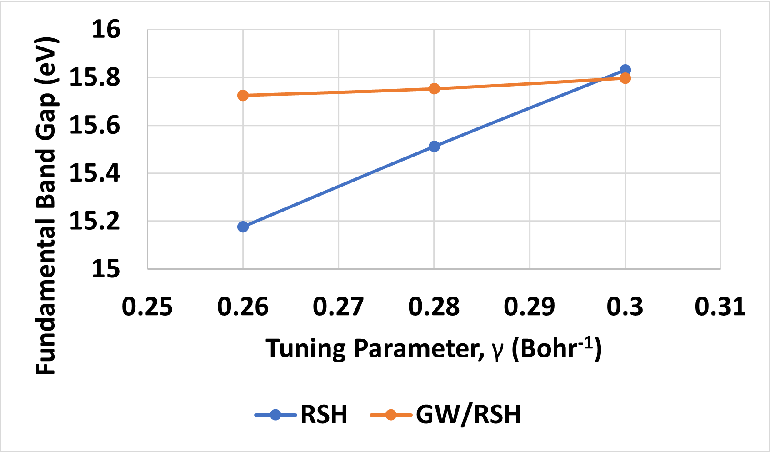} 
\par\end{centering}
\caption{Fundamental band gaps (in eV) as functions of $\gamma$ for a LiF
$5\times5\times5$ supercell.}
\end{figure}

It is evident from Fig. 1 that the RSH DFT results are more sensitive
to $\gamma$ than the GW gaps. This is expected, as $\gamma$ affects
the GW gap only indirectly by changing the DFT starting point. A summary
of the calculated fundamental band gaps are given in Table 1.

\begin{table}
\begin{centering}
\begin{tabular}{|c|c|c|c|c|}
\hline 
 & \multicolumn{2}{c|}{OT-GW/RSH} & \multirow{1}{*}{scGW} & \multirow{1}{*}{Exp.}\tabularnewline
\hline 
 & VASP  & sGW  &  & \tabularnewline
\hline 
\hline 
ZnO  & 3.7  &  & 3.8$^{a}$  & 3.44$^{a}$\tabularnewline
\hline 
Si  & 1.24  & 1.24  & 1.41$^{a}$  & 1.17$^{a}$\tabularnewline
\hline 
LiF  & 15.9  & 15.8  & 15.9$^{a}$  & 14.2$^{a}$\tabularnewline
\hline 
CdO  & 0.99  &  & 0.98$^{b}$  & 0.84$^{b}$\tabularnewline
\hline 
\end{tabular}
\par\end{centering}
\caption{Fundamental band gaps with OT-GW/RSH using VASP and sGW, in eV. Results
from self-consistent GW (scGW) as well as experiments (Exp.) are also
reported. References: a)\cite{shishkin2007accurate}b)\cite{deguchi2016accurate}}
\end{table}

With the exception of Si, results obtained using OT-GW/RSH are in
good agreement with that of self-consistent GW (scGW). This is consistent
with our notion that OT-GW/RSH is equivalent to finding a stationary
point in scGW. We also note that both OT-GW/RSH and scGW tend to over-estimate
the experimental value. In fact, this over-estimation has been reported
in the literature, and this performance has been ascribed to an under-estimation
of the dielectric screening in random-phase approximation (RPA) adopted
in GW calculations. \cite{van2006quasiparticle,shishkin2007self}
Possible ways to fix the situation were proposed,\cite{chen2015accurate,shishkin2007accurate}
and this will be explored in our future work.

As for the present work, we emphasize that OT-GW/RSH provides a way
of tuning the range-separation parameter for periodic systems, and
it produces good results. To see this, the fitted optimal $\gamma$
are reported in Table 2. For reference, we compare the optimal $\gamma$
obtained from OT-GW/RSH to that calculated using the empirical formula
of Baer et al.\cite{Baer2009} The two methods give results that are
overall consistent though not identical. This indicates that OT-GW/RSH
is an effective method for obtaining the range-separation parameter
in periodic systems from first principles only.

\begin{table}
\begin{centering}
\begin{tabular}{|c|c|c|c|c|}
\hline 
 & $\gamma$(VASP)  & $\gamma$(sGW)  & $\gamma$(Fitted)  & $\epsilon_{\infty}$\tabularnewline
\hline 
\hline 
ZnO  & 0.12  &  & 0.09  & 3.14$^{a}$\tabularnewline
\hline 
Si  & 0.029  & 0.029  & 0.019  & 11.68$^{b}$\tabularnewline
\hline 
LiF  & 0.286  & 0.297  & 0.2  & 1.92$^{c}$\tabularnewline
\hline 
CdO  & 0.1  &  & 0.15  & 2.3$^{d}$\tabularnewline
\hline 
\end{tabular}
\par\end{centering}
\caption{Optimally tuned $\gamma$ in Bohr$^{-1}$, obtained from OT-GW/RSH
with VASP and sGW. We also report values fitted from dielectric constant
using the empirical formula from the work of Baer et al.\cite{Baer2009}
References: a)\cite{zno2013dielectric}b)\cite{Baer2009}c)\cite{lif1972digest}d)\cite{cdo2015372}}
\end{table}

\section*{Acknowledgements}

The authors acknowledge support from the Center for Computational
Study of Excited State Phenomena in Energy Materials (C2SEPEM) at
the Lawrence Berkeley National Laboratory, which is funded by the
U.S. Department of Energy, Office of Science, Basic Energy Sciences,
Materials Sciences and Engineering Division under Contract No. DE-AC02-05CH11231
as part of the Computational Materials Sciences Program. Computational
resources were supplied through the XSEDE allocation TG-CHE170058.
In addition, R.B. gratefully acknowledge the support from the US\textminus Israel
Binational Science Foundation (BSF) under Grant No. 2018368.

\bibliographystyle{unsrt}
\bibliography{ref}

\end{document}